\documentclass[12pt]{iopart}

\usepackage{units}
\usepackage{graphicx}
\usepackage{bm}
\usepackage{color}

\begin{document}

\title{MnSi-nanostructures obtained from thin films: magnetotransport and Hall effect}

\author{D Schroeter$^{1}$, N Steinki$^{1}$, M Schilling$^{1}$, A Fern\'andez Scarioni$^{2}$, P Krzysteczko$^{2}$, T Dziomba$^{2}$, H W Schumacher$^{2}$, D Menzel$^{1}$ and S S\"ullow$^{1}$}

\address{$^{1}$Institut f\"ur Physik der Kondensierten Materie, Technische Universit\"at Braunschweig, D-38106 Braunschweig, Germany\\}
\address{$^{2}$Physikalisch-Technische Bundesanstalt, D-38116 Braunschweig, Germany}

\date{\today}

\begin{abstract}
We present a comparative study of the (magneto)transport properties, including Hall effect, of bulk, thin film and nanostructured MnSi. In order to set our results in relation to published data we extensively characterize our materials, this way establishing a comparatively good sample quality. Our analysis reveals that in particular for thin film and nanostructured material, there are extrinsic and intrinsic contributions to the electronic transport properties, which by modeling the data we separate out. Finally, we discuss our Hall effect data of nanostructured MnSi under consideration of the extrinsic contributions and with respect to the question of the detection of a topological Hall effect in a skyrmionic phase.
\end{abstract}


\maketitle


\section{Introduction}

In recent years, {\it skyrmions}, {\it i.e.}, a unique form of complex magnetic spin texture, have evolved as a research topic of prime interest both for basic research as from the perspective of possible applications \cite{muehlbauer2009,krause2016}. With respect to the latter issue, heterostructures carrying skyrmionic spin textures hold the prospect of being used for new types of data storage devices \cite{fert2013,tomasello2014}. Conversely, the former aspect bears relevance in the context of a multitude of fundamental topics such as topologically protected states, spin-orbit effects in magnetic materials or the collective dynamics of the skyrmionic phase \cite{neubauer2009,schuette2014,schwarze2015}.

The basic structural element required to stabilize skyrmions is a lack of inversion symmetry in the magnetic lattice, leading to a spin-orbit controlled type of magnetic exchange, the Dzyaloshinskii-Moriya (DM) interaction. For two spins $S_i$, $S_j$, in contrast to the ordinary Heisenberg exchange, the DM interaction results in an energetic term $\propto S_i \times S_j$. If combined with the Heisenberg exchange, and in a Ginzburg-Landau type approach, it has been demonstrated that a skyrmionic phase can be stabilized under these conditions \cite{muehlbauer2009}. In turn, with respect to materials exhibiting skyrmionic phases, the occurrence of a DM interaction can be attributed either to crystal structures lacking inversion symmetry (as for instance in the B20 compounds \cite{pfleiderer2010}), or to the symmetry-breaking at surfaces \cite{bode2007,heinze2011}. Correspondingly, in terms of the appearance of skyrmionic phases these are either surface induced thin film or material intrinsic skyrmions that are studied. 

Belonging to the latter class, the cubic helimagnetic B20 compound MnSi represents probably the most iconic system exhibiting a skyrmionic phase. The material being known for decades (see Ref.~\cite{pfleiderer2010} for a review) was originally studied in the context of spin fluctuation theory \cite{moriya1985}. Later, the pressure induced suppression of helical magnetic order (helix length $\sim 19$\,nm, ordered magnetic moment $\sim 0.4 \mu_B$ at ambient pressure) became the focus of studies in the context of quantum criticality in itinerant $d$-metals \cite{pfleiderer2010,thompson1989}. Finally, it was noted that the early-reported field-induced {\it A-}phase in MnSi \cite{sakakibara1982} does represent a skyrmion phase \cite{muehlbauer2009}, this way establishing the material as the model compound for studies of skyrmion physics.

Subsequently, one avenue that was followed in investigations on MnSi was the production of thin film material \cite{karhu2010,karhu2011,karhu2012,geisler2012,wilson2012,engelke2012,li2013,suzuki2013,menzel2013,wilson2013,yokouchi2014,wilson2014,engelke2014,meynell2014a,meynell2014b,yokouchi2015,lancaster2016}. Conceptually, it was argued that in thin films the skyrmionic phase ought to be stabilized \cite{butenko2010,banerjee2014,rowland2016}, and thus it should be possible to better control and investigate skyrmionic behavior. However, in spite of these extensive studies, as yet, there is no consensus regarding the issue if in thin film material of MnSi a skyrmionic phase exists in a similar form as in bulk material.

Initial studies established that MnSi thin films do show helical magnetic order, although for films thicker than $\sim 5$\,nm at temperatures significantly higher ($T_C$ up to $\sim 48$\,K) than in bulk material ($T_C = 29.5$\,K). This effect is attributed to tensile strain induced in MnSi thin films by the lattice mismatch with the Si substrate, effectively leading to a state of negative pressure \cite{karhu2010,engelke2012,engelke2014}. As well, the suppression of the ordering temperature for films with less than 5\,nm thickness can be understood in terms of the reduction of spin-spin interactions caused by the interface \cite{engelke2012}. 

For films thicker than the helix length of bulk material there is evidence that the magnetic ground state is essentially the same as in the bulk, be it that uniaxial anisotropies induced by thin film strain need to be taken into account \cite{karhu2010,karhu2011,karhu2012,engelke2012,li2013,wilson2013,engelke2014}. For the in-field behavior, however, it turned out that the film anisotropies affect the material properties quite significantly, leading to a variety of proposed magnetic phase diagrams \cite{wilson2012,li2013,menzel2013,yokouchi2014,lancaster2016}. Most notably, while for single crystals MnSi that have been mechanically thinned to a few 10\,nm thickness, enabling studies by Lorentz microscopy \cite{tonomura2012,mochizuki2014,yu2015}, there is direct evidence for a skyrmionic in-field phase, as yet there is no conclusive evidence for such a phase in MnSi thin films.

In this situation, an attempt has been made to identify the skyrmion phase in MnSi thin films by a unique feature of the skyrmionic state, the {\it topological Hall effect} \cite{neubauer2009,li2013,yokouchi2014,meynell2014b,yokouchi2015}. Experimentally, it requires accurate measurements of the Hall effect of the system studied, as the topological Hall effect in bulk material represents a minor additional Hall contribution aside from normal and anomalous Hall effect. Surprisingly, even though the studies on thin film material in the Refs.~\cite{li2013,yokouchi2014,meynell2014b,yokouchi2015} appear to verify that in terms of the general magnetic behavior the different MnSi thin films behave in a similar way, the measurements of the Hall effect in these studies exhibit widely varying experimental features. 

As a reference, the analysis to extract the topological Hall effect in bulk MnSi has been performed on data taken for high quality single crystalline plates of about 100\,$\mu$m thickness (room temperature resistivity $\rho_{xx} = 180$\,$\mu \Omega$cm, RRR\,$\sim$\,100) \cite{neubauer2009}. The Hall effect and magnetoresistivity have been measured in a standard 6-point configuration up to 9\,T, with the signal being symmetric/antisymmetric in a magnetic field corresponding to the magnetoresistivity/Hall effect. Overall, the magnitude of the Hall signal $\rho_{yx}$ varies between -150 and 200\,n$\Omega$cm, and its field dependence is dominated by the normal contribution and an anomalous contribution reflecting primarily the magnetization of the sample. After correction for the normal Hall contribution (carrier density $\sim 4 \times 10^{22}$\,cm$^{-3}$) an anomaly in the field dependence of the anomalous Hall contribution is observed. This anomaly, with a magnitude of about 5\,n$\Omega$cm on the background of $\sim 50$\,n$\Omega$cm for the ordinary anomalous Hall contribution, has been demonstrated to map the {\it A-}phase of MnSi.

In comparison, in Table \ref{tab:comp} we summarize basic physical parameters for different MnSi thin films reported in literature. As noted before, for a film thickness above 5\,nm an enhanced ordering temperature $\sim$\,45\,K is observed. However, already the absolute values of the resistivity vary significantly, even though they are in the range of single crystalline material. Clearly, disorder in the films is an issue, as exemplified by residual resistivity ratios significantly smaller than for the best single crystals, and varying from sample to sample. As well, sample dependencies become very apparent if the Hall resistivities are considered. As indicated by the values reported at 20\,K in 1 and 5\,T, for the different samples they vary in magnitude and even by sign.

\begin{table}
\begin{tabular}{|c|c|c|c|c|c|c|} \hline
Ref. & \cite{engelke2012} & \cite{li2013} & \cite{li2013} & \cite{yokouchi2014} & \cite{meynell2014b} & \cite{yokouchi2015} \\ \hline
thickness (nm) & 19 & 10 & 50 & 20 & 25.4 & 26 \\
$T_C$ (K) & 45 & 45 & 45 & 45 & 42 & 48 \\
$\rho_{xx: 80 \textrm{K}} (\mu \Omega \textrm{cm}$) & 99 & 165 & -- & 120 & 113 & -- \\ 
$\rho_{xx: 80 \textrm{K}} / \rho_{xx: 2 \textrm{K}}$ & 11 & 7 & -- & 12 & 14 & -- \\
$\rho_{yx: 20 \textrm{K}/1 \textrm{T}} (\textrm{n} \Omega \textrm{cm}$) & -- & 5 & -10 & -10 & 10 & -- \\
$\rho_{yx: 20 \textrm{K}/5 \textrm{T}} (\textrm{n} \Omega \textrm{cm}$) & -- & 65 & 60 & 40 & 50 & -- \\
$n (10^{22}$\,cm$^{-3}$) & -- & 3.5 & 3.5 & 8.8 & 4.3 & -- \\
\hline
\end{tabular}
\caption{List of main physical characteristics of MnSi thin film samples reported in the literature: ordering temperature $T_C$, resistivity $\rho_{xx: 80 \textrm{K}}$ at 80\,K, resistivity ratio $\rho_{xx: 80 \textrm{K}} / \rho_{xx: 2 \textrm{K}}$, Hall resistivity $\rho_{yx: 20 \textrm{K}/x \textrm{T}}$ in 1 and 5\,T, carrier density $n$ extracted from the normal Hall contribution.}
\label{tab:comp}
\end{table}

To extract the topological contribution to the Hall effect from the experimental data, the conceptual idea is to separate the Hall resistivity into three parts:
\begin{equation}
\rho_{yx} = \rho_{yx}^N + \rho_{yx}^A + \rho_{yx}^T \label{Hall}
\end{equation}
Here, $\rho_{yx}^N$ denotes the normal Hall contribution ({\it i.e.}, the carrier density dependent term), $\rho_{yx}^A$ the anomalous part proportional to the magnetization and a resistivity-dependent factor, while $\rho_{yx}^T$ represents the topological Hall effect. Given that $\rho_{yx}^N$ is roughly the same for all samples and varies linearily with field, the difficulties in extracting a topological Hall contribution is directly related to the accurate determination of $\rho_{yx}^A$. From the experimental data, however, it is apparent that $\rho_{yx}^A + \rho_{yx}^T$ is not well-controlled, a point that has been made previously in an attempt to more accurately parametrize the Hall resistivity \cite{meynell2014b}.

In this situation, we have set out to reinvestigate the Hall effect in MnSi thin films. We do so by 1.) carefully comparing bulk and thin film data and 2.) nanostructuring MnSi thin films into Hall bar geometries for optimization of the experimental set-up. We analyze the electronic transport properties of different nanostructures MnSi, this way assessing which contributions to the resistive and Hall behavior of thin film material are intrinsic or extrinsic, respectively. 

The paper is organized as follows: First, we report the sample preparation and experimental analysis to characterize our MnSi thin films, this in order to compare our samples to thin film samples from previous reports. Next, we describe our steps to nanostructure the thin films and document our resulting Hall bar structures. Subsequently, we characterize our structures regarding their electronic transport properties, {\it i.e.}, the resistivity, magnetoresistivity and Hall effect. Finally, we discuss our findings on nanostructured MnSi, this in particular in comparison to thin film and single crystalline material and with respect to the issue of the existence of a skyrmion phase in these samples.

\section{Sample preparation}
\label{sample}

MnSi thin film samples were prepared by molecular beam epitaxy (MBE) in ultra-high vacuum with a base pressure of below 5$\times$10$^{-11}$ mbar. Essentially, the procedure has been described previously \cite{engelke2012}, but was carried out here with some alterations. $P$-doped Si(111) wafers with a size of $10 \times 10 \times 0.28$\,mm$^3$ and a resistivity of $1 - 10$\,$\Omega$cm serve as substrates. After a standard cleaning process the substrates were loaded into the MBE chamber, where they were degassed at 750$^\circ$C for two hours. 

To remove the native silicon-oxide layer the temperature was then raised at a rate of 1$^\circ$C$/$s up to 1150$^\circ$C were the system was held for 10 min after which the substrate was cooled back to room temperature at a rate of less than 1$^\circ$C$/$s. After this processing step a $7 \times 7$ reconstruction of the Si surface can be clearly seen in reflection high energy electron diffraction (RHEED), confirming the high crystalline quality of the substrate. Subsequently a seed layer for the MnSi films was formed by evaporating 1\,nm Mn from a Knudsen cell onto the substrate, which is held at a temperature of 180$^\circ$C. As the next step, the substrate is heated to 300$^\circ$C, at which an epitaxial MnSi seed layer is formed. In our experiments, a series of different Mn layer thicknesses between 0.5 and 4\,nm were tested, with a layer thickness of 1\,nm showing the clearest $\sqrt{3} \times \sqrt{3}$ pattern visible in RHEED.

Following the growth of the seed layer, MnSi thin films of various thicknesses were grown by simultaneous deposition of Mn and Si in a stoichiometric ratio, with Si evaporated using an electron beam evaporator. The overall deposition rate was 0.02\,nm$/$s. After growing the MnSi layer the sample is annealed at 350$^\circ$C for 1 hour, always showing the characteristic $\sqrt{3} \times \sqrt{3}$ RHEED pattern. The single phase nature of the samples annealed at 350$^\circ$C were checked by x-ray diffraction, and revealed no detectable impurity phase. As reported in Ref. \cite{karhu2010}, we observe that annealing at higher temperatures than 400$^\circ$C tends to produce MnSi$_{1.7}$ precipitates. Altogether, following above recipe we produced films of a nominal thickness of 10 and 30\,nm. Below, we present measurements of the electronic transport properties of the 30\,nm thick samples.

\section{Surface analysis}
\label{surface}

In our earlier studies on the (magneto)transport and magnetization of MnSi thin films \cite{engelke2012,menzel2013,engelke2014} we have noticed that the smoothness of the surface is decisive for the sample quality. So far, no roughness data of the film surface has yet been published. Earlier investigations on film morphology have probed only the MnSi/Si interface via x-ray reflectometry, and which has been reported to be of the order of 1\,nm \cite{karhu2010,karhu2011}. Therefore, in order to assess the quality of our thin film samples, we have carried out atomic force microscopy (AFM) measurements on various of our 30\,nm thick samples with a SIS Nanostation II non-contact AFM system (by Surface Imaging Systems SIS, Germany; now known as N8 NEOS by Bruker), using a PPP-NCLR cantilever (Nanosensors) with scans ranging from 100\,$\mu$m\,$\times$\,100\,$\mu$m down to 3\,$\mu$m\,$\times$\,3\,$\mu$m recorded with 1024\,$\times$\,1024 data points each.

\begin{figure}
\centering
		\includegraphics[width=1 \columnwidth]{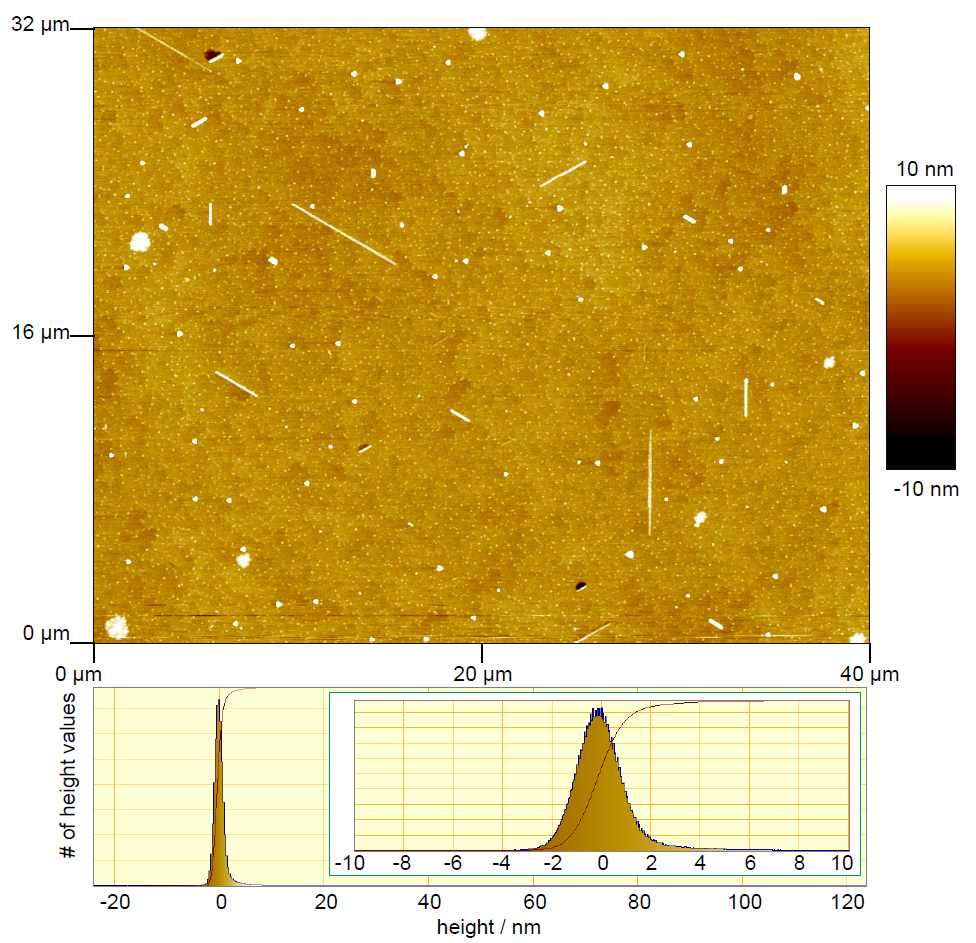}
	\caption{(Color online) AFM image and height histogram of a MnSi film grown on Si(111) with nominal thickness of 30\,nm; for details see text.}
	\label{afmglobal}
\end{figure}

\begin{figure}
\centering
		\includegraphics[width=1 \columnwidth]{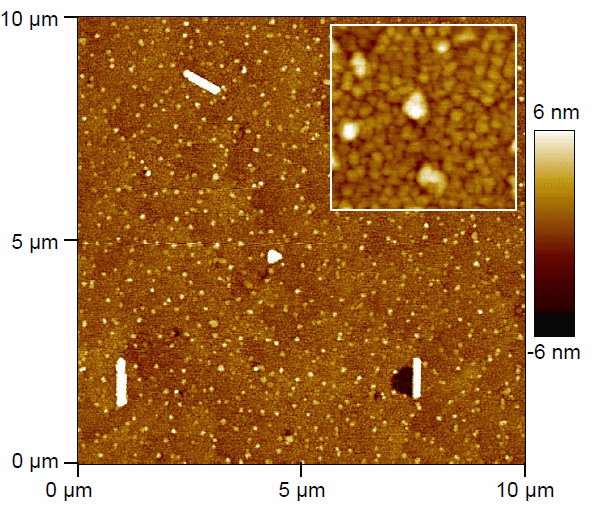}
	\caption{(Color online) AFM image of a MnSi film grown on Si(111), with a 10\,$\mu$m\,$\times$\,10\,$\mu$m area depicted, and a zoom-in to a 1\,$\mu$m\,$\times$\,1\,$\mu$m area in the inset; for details see text.}
	\label{afmzoom}
\end{figure}

In the Figs. \ref{afmglobal} and \ref{afmzoom} we display the results of AFM measurements on two nominally 30\,nm thick films. Overall, the AFM images show a flat closed MnSi layer with a very uniform surface, with Sq values (RMS roughness) well below 3\,nm. A major contribution to the roughness is made by needle-like features of 10 to 20\,nm height arranged in three discrete orientations (see white needles in Figs. \ref{afmglobal} and \ref{afmzoom}), and some grains (larger white dots in Figs. \ref{afmglobal} and \ref{afmzoom}) mainly of some ten nanometers height. Needles and grains occupy less then 1\,\% of the surface, and the remaining 99\,\% of the surface shows Sq values of about 1\,nm only. Considering that the film surface roughness increases during the growth process, the low roughness measured here indicates a good film quality. 

Throughout the AFM measurements, also the phase signal of the AFM was monitored in order to check for contamination with other materials, but the phase images show very small contrasts only. These should be attributed to topographic features rather than secondary phase materials. In particular, even the pronounced white structures (needles and grains in Fig. \ref{afmglobal}) show only topological features in the phase signal, indicating that in fact we are dealing with a single phase MnSi film. Conversely, considering the well defined topology of the needles and grains, we assume that these are whisker crystals MnSi growing out of the thin films. 

\section{Lithography}
\label{lithography}

One experimental issue while measuring the Hall effect is to place the Hall voltage contacts directly opposite to each other on the sample. In reality, imperfections in this step always lead to a magnetoresistive component in the measured signal, effectively reducing the signal-to-noise ratio for the intrinsic Hall signal. Therefore, in order to minimize the magnetoresistive contribution, and in perspective to allow studies on the controlled movement of skyrmions, we have set out to nanostructure our MnSi thin film material by means of electron beam lithography. In a first step, we have produced Hall bar structures, allowing simultaneously to measure the resistivity, magnetoresistivity and Hall effect on our samples.

\begin{figure}
\centering
		\includegraphics[width=1 \columnwidth]{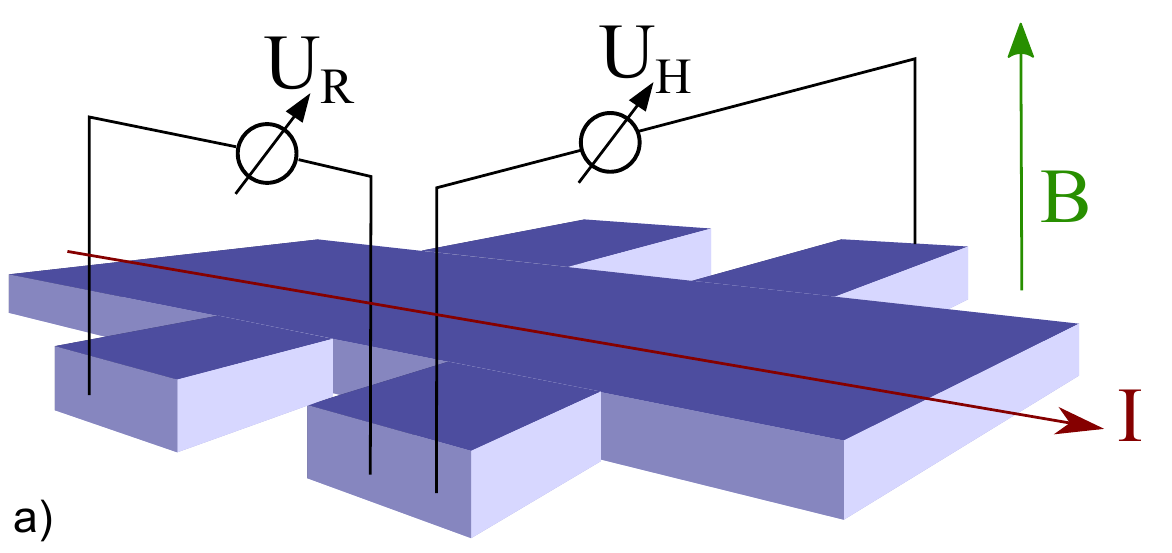}
		\includegraphics[width=1 \columnwidth]{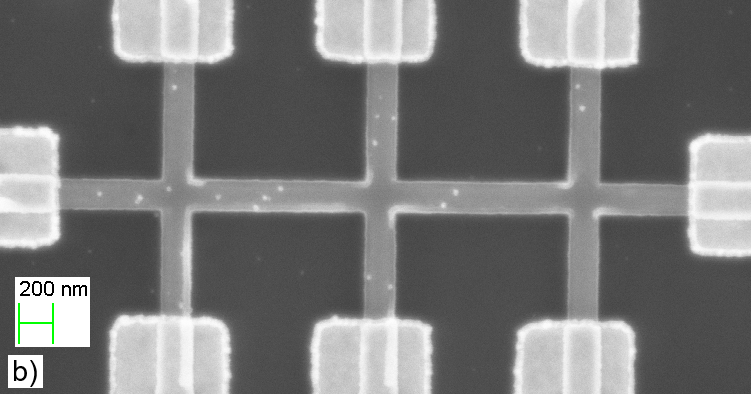}
	\caption{(Color online) (a) Schematic view of the microfabricated Hall bar structures, including a sketch of experimental geometries. (b) Scanning electron microscope image of a MnSi nanostructure produced by electron beam lithography with a current path width of 175\,nm; for details see text.}
	\label{rem}
\end{figure}

Standard electron beam lithography in combination with Ar-ion etching was used to pattern the 30\,nm thick MnSi films into Hall bar geometry with a measured width (by means of scanning electron microscopy (SEM), see below) of the current path ranging from 175\,nm to 10\,$\mu$m. The basic design for most of our samples consists of three crossings (Fig. \ref{rem} (b)). The samples have been prepared by using the electron beam resist \mbox{AR-N 7520} with an additional precleaning step by sonicating the sample in \mbox{mr-Rem 660}, which significantly improves the adhesion between the sample and the resist.

In the next lithography step platinum (Pt) contacts to the MnSi structure were fabricated by sputtering deposition. The Pt leads are 100\,nm thick, including a 5\,nm Ta adhesion layer. The MnSi surface was cleaned in situ beforehand by low energy (0.7\,keV) Ar-ions to ensure good electrical contact. This step does not influence the magnetic properties as verified by magnetization measurements (not shown). Moreover, simulations of Ar-ion implanting in our films indicate that the low-energy Ar-ions will penetrate the surface by roughly 1\,nm. We thus conclude that the Ar-etching does not affect the properties of our structures.   

As results of these different processing steps we have obtained various Hall bar structures from MnSi thin films. For final characterization, we have carried out scanning electron microscopy on all nanostructures reported on here, with the example of the 175\,nm structure depicted in Fig. \ref{rem}(b). The structure contact pads were bonded with aluminum wires to the sample carrier, which was attached to the sample holder of our cryogenic system for measurements of (magneto)resistivity and Hall effect.

\section{Experimental}
\label{experimental}

To characterize the MnSi structures we have measured the temperature and field dependent electronic transport properties in a $^4$He cryostat via standard {\it ac}, {\it dc} four-point probe and van der Pauw configuration in the temperature range 2\,--\,300\,K in fields up to 5\,T. The magnetic field was applied perpendicular to the film surface (see Fig. \ref{rem}(a)). 

We have carried out these experiments for different structure sizes with a current path width between 10\,$\mu$m and 175\,nm. Effectively, with the {\it ac} excitation voltage from a low-power resistance bridge, we use measurement currents in the range of 10\,nA up to 1\,$\mu$A, which corresponds to current densities of $1\cdot 10^{6}\,$A/m$^2$ for $\mu$m sized structures and up to $2\cdot10^{7}\,$A/m$^2$ for the 175\,nm structures. For the {\it dc} measurements (performed on the 200\,nm and 230\,nm structures) a current of 100\,$\mu$A was used which corresponds to current densities of roughly $2\cdot10^{10}\,$A/m$^2$. In terms of skyrmionic behavior, if skyrmions are present in our samples, already the {\it ac} currents could possibly be large enough to electrically drive such skyrmions through our structures \cite{jonietz2010}. 
 
To directly compare data of MnSi single crystals, thin films and patterned structures, in addition to the nanostructured material, we have measured on a 30\,nm MnSi thin film without lithographical treatment the temperature dependent (magneto)resistivity (current density of $2\cdot 10^{6}\, $A/m$^2$) and Hall effect (current density of $4\cdot 10^{7}\, $A/m$^2$). Furthermore we have determined the corresponding properties of a 2\,mm\,$\times$\,1\,mm\,$\times$\,0.1\,mm single crystalline sample which was synthesized by tri-arc Czochralski growth. 

Geometrical factors required to obtain absolute resistivity values for our nanostructures were determined by the analysis of the scanning electron microscopy pictures. This way, the length $l$ probed in the resistance measurements of our nanostructured samples can be determined easily to an accuracy of about one percent. As well, the width of the current path is a well-controlled quantity, with an uncertainty (depending on structure size) of a few percent. The largest error margin comes from the film thickness, with an uncertainty of about 10\,\% between nominal and real thickness. Only, this error does not affect our comparison of thin film and nanostructured samples, as the latter ones are produced from the films. Altogether, the geometry induced uncertainty of absolute resistivity values is of the order of 10\,\%, the relative error margin in sample-to-sample comparison in the percentage range.  

\section{Results}
\label{results}

\subsection{Resistivity}
\label{elcharacter}

\begin{figure}
\centering
		\includegraphics[width=1 \columnwidth]{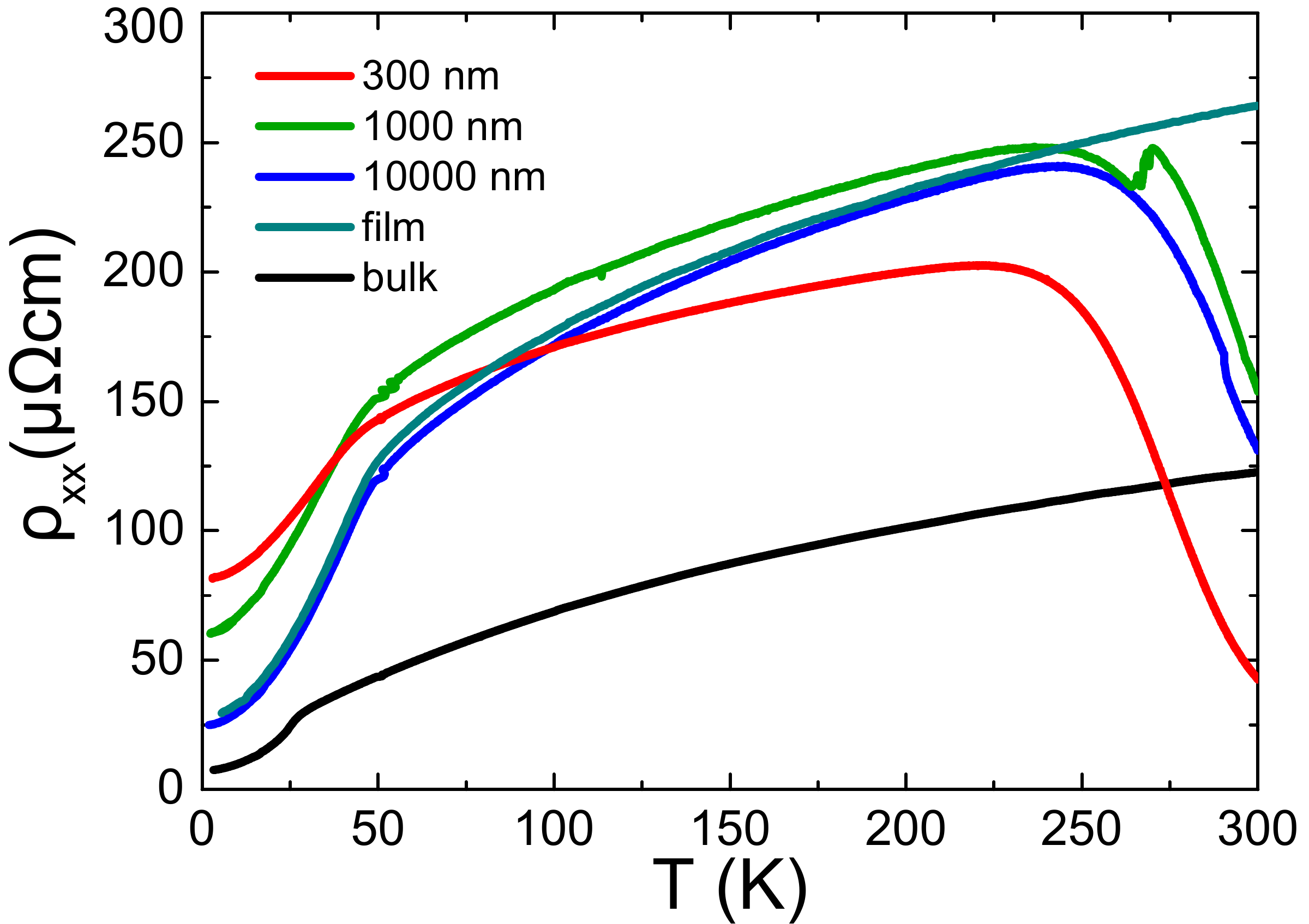}
	\caption{(Color online) Temperature dependence of the resistivity $\rho_{xx}$ of bulk, thin film and nanostructured thin film MnSi (film thickness 30\,nm); for details see text.}
	\label{substrate}
\end{figure}

In Fig. \ref{substrate} we summarize the resistivities of different MnSi samples, ranging from bulk material to nanostructured thin films (thickness 30\,nm). For bulk material, we reproduce the essential findings from literature \cite{pfleiderer2010,kadowaki1982}, that is a resistivity from an itinerant $d$-metal magnetic system, with a room temperature resistivity of about 120\,$\mu \Omega$cm (Fig. \ref{substrate}). A residual resistivity ratio $\rho_{xx: 300 \textrm{K}} / \rho_{xx: 2 \textrm{K}}$ of 16 signals a decent crystalline quality of our specimen.

For a thin film sample that was contacted by Cu wires with silver epoxy to the surface of the film, again we reproduce the resistivity reported previously \cite{engelke2012,li2013,meynell2014b}. Overall, there is a metallic resistivity with a kink-like feature denoting the magnetic transition at $T_C$. A residual resistivity $\rho_{xx: 2 \textrm{K}} = 20 \mu \Omega$cm and a resistivity ratio $\rho_{xx: 80 \textrm{K}} / \rho_{xx: 2 \textrm{K}} = 8$ fully agrees with literature values.

For thin films nanostructured in Hall bar geometry on $P$-doped Si(111)-substrates, in addition to the resistive behavior typical for MnSi film there is a downturn of $\rho_{xx}$ at high temperatures (Fig. \ref{substrate}). If we specifically consider the largest structures, say the structure with a lateral extent of 10\,$\mu$m structure width and 30\,$\mu$m structure length, the obvious expectation is that at this structure size there should be no size effect, {\it i.e.}, the behavior of the MnSi structure and the film should be identical. Indeed, the direct comparison of the data for a 10\,$\mu$m structure and the thin film reveals the resistivity of film and structure to be almost the same up to 230\,K. But then, the downturn at higher $T$ must reflect the influence of the substrate and the contact pads on the measurement, effectively providing a resistive path parallel to the MnSi structure. 

\begin{figure}
\centering
		\includegraphics[width=1 \columnwidth]{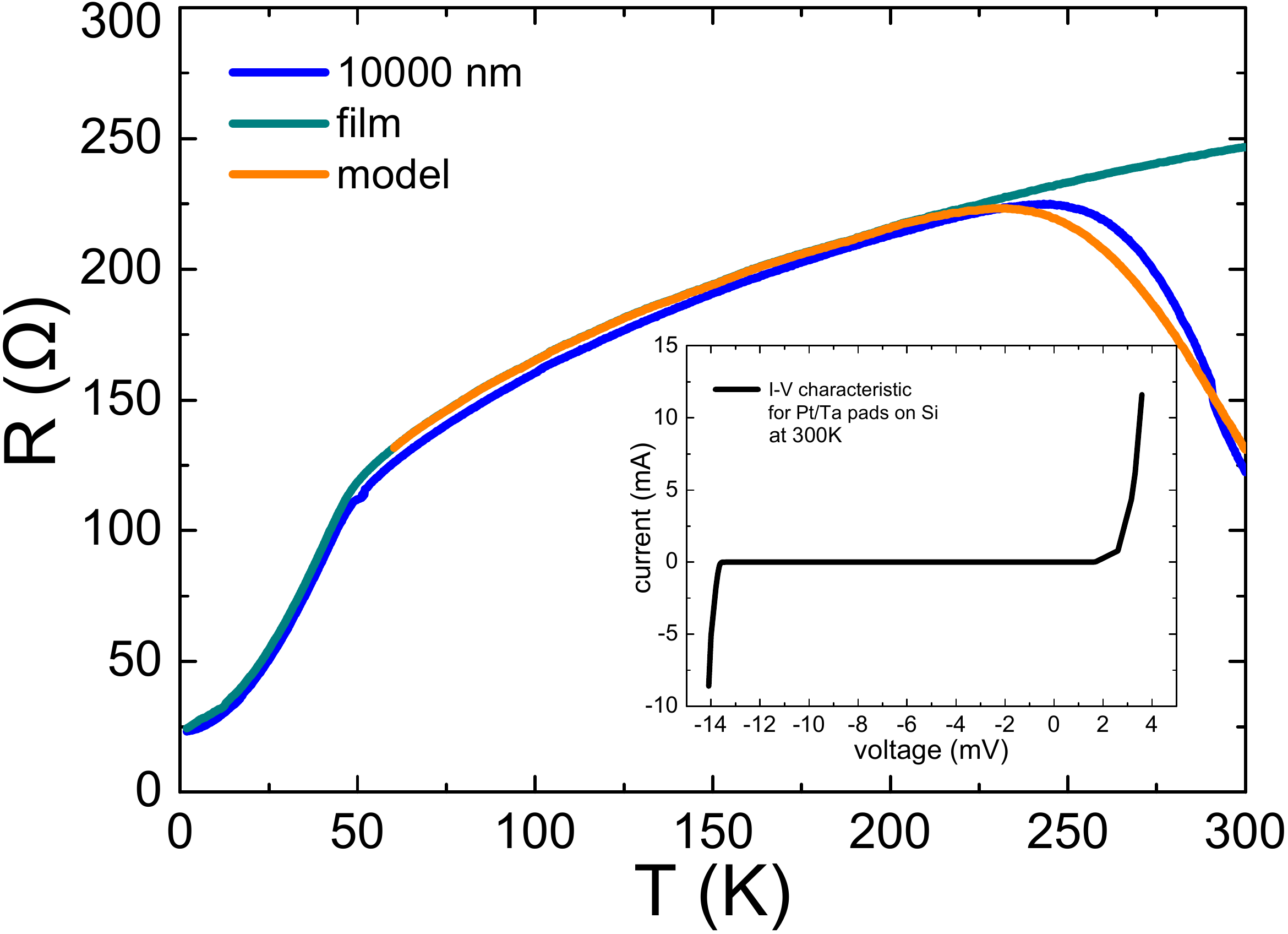}
	\caption{(Color online) Temperature dependence of the resistance $R$ of MnSi film and patterned sample (film thickness 30\,nm, structured path width 10\,$\mu$m) on Si-substrate: a downturn in the resistivity of the structured samples can be seen as a result of a parallel resistor model (orange) between MnSi film and substrate, for details see text. Inset: $I-V$ characteristic measured at 300\,K of the contact pads only on the Si-substrate displaying a Schottky barrier behavior.}
	\label{modeling}
\end{figure}

To demonstrate this, we have modeled our measured signal, assuming that the total resistance $R_{total}$ of nanostructured MnSi (current path width 10\,$\mu$m) on Si-substrate can be understood as a set of parallel resistors (Fig. \ref{modeling}):
\begin{equation} 
R_{total} = \frac{R_{MnSi} \cdot R_{sub}}{R_{MnSi} + R_{sub}}.
\end{equation} 
Here, we use the total resistance of the substrate $R_{sub}$ which was measured using the contact pads of two different and separated MnSi nanostructures, displaying an exponential temperature dependence. In addition, we use the temperature dependent part of the experimentally determined resistivity $\rho_{exp}(T)$ of thin film MnSi to calculate $R_{MnSi} = (l/a) \rho_{exp}(T)$ for the nanostructure. Here, we use the MnSi structure length $l$ and the cross-section $a$. In Fig. \ref{modeling} we demonstrate that with these simple assumptions we can model the temperature dependence of the resistance of a Si-substrate chip carrying a Hall-bar structure MnSi of 10\,$\mu$m width by comparing the experimental data with the calculated resistance following the parallel resistor approach. It proves that because of the strong temperature dependence of the resistance of the substrate, at high temperatures of roughly $T>$\,200\,K the resistivity of the substrate increasingly affects the overall measurement. 

The resistance $R_{sub}$, which exponentially increases with decreasing temperature, does not reflect the intrinsic resistive behavior of a $P$-doped Si(111) wafer. Instead, it appears that the resistive contribution short circuiting the MnSi structure at high temperatures is related to a Schottky barrier at the interface between the contact pads and the Si wafer. This is indicated by the current-voltage characteristic of the contact pads connected to MnSi nanostructures (see inset of Fig.\,\ref{modeling}). For the measurement plotted here (at a temperature of 300\,K), two pads on one MnSi structure were used as $I$ and $V$ contacts, while for the second $I/V$ contacts the pads on a second MnSi structure on the same wafer were used (distance between structures $\sim 80$\,$\mu$m). In this configuration we thus probe the resistance through the contacts and substrate. The measured $I-V$ characteristic is typical for a Schottky barrier. Moreover, as we lower the temperature, the non-conducting voltage range and the resistance increases, reflecting an increasingly insulating resistance path through the wafer. 

Effectively, our finding implies that for the nanostructured samples MnSi there are Schottky barriers between the contact pads and the Si wafer. At high temperatures, these short-cut the nanostructures, while at lower temperatures the barrier becomes impassable, in result decoupling the MnSi nanostructures from the substrate. The quantitative agreement between experimental data and our parallel resistor model proves that at high temperatures (above $\sim 200$\,K) the measured resistance of the nanostructures is significantly affected by the substrate/contact pads. Conversely, with the exponentially increasing resistivity of the substrate/contact pads for lower temperatures, in the temperature range considered below, $T < 80$\,K, an influence of the substrate on $\rho_{xx}(T)$ of MnSi can be neglected.

\begin{figure}
\centering
		\includegraphics[width=1 \columnwidth]{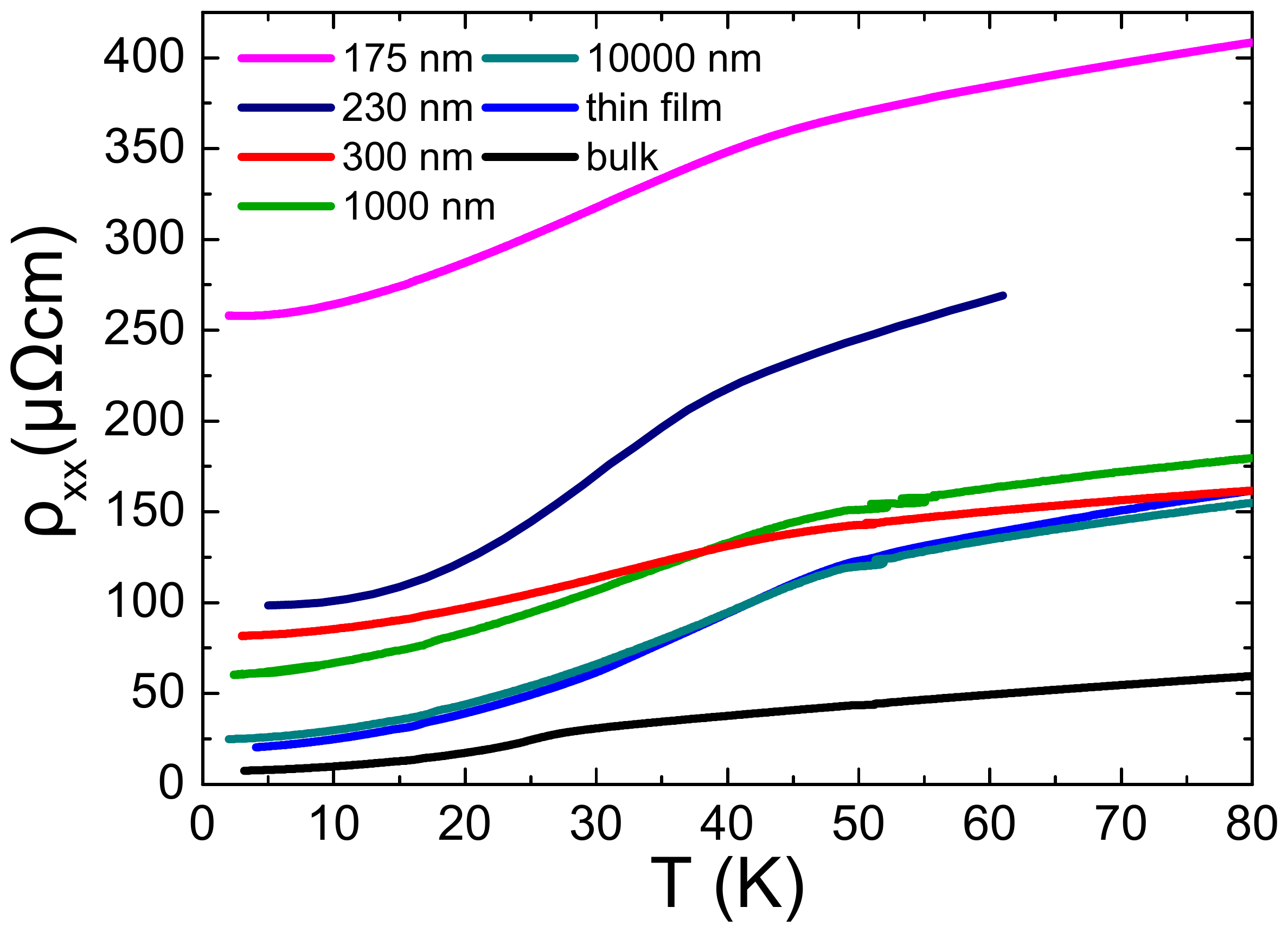}
	\caption{(Color online) Resistivity of different MnSi samples (bulk, thin film and nanostructures) in zero magnetic field up to 80\,K; for details see text \cite{note2}.}
	\label{resistivity}
\end{figure}

Having thus identified the different contributions to the resistive behavior, in Fig. \ref{resistivity} we summarize the resistivities of different samples MnSi, ranging from bulk material and unpatterned films to nanostructured films, at temperatures $T < 80$\,K. In addition, in Tab. \ref{tab:samples} we list the essential parameters obtained from these measurements, {\it i.e.}, transition temperatures, resistive coefficients, residual resistivities and resistivity ratios. 

\begin{table}
\footnotesize
\begin{tabular}{|c|c|c|c|c|c|c|c|} \hline
Sample & bulk & film & 10\,$\mu$m & 1\,$\mu$m & 300\,nm & 230\,nm & 175\,nm \\ \hline
$T_C$ (K) & 29.5 & 48 & 48 & 47 & 45 & 40 & 44\\
$A$ ($\mu \Omega$cm/K$^2)$ & 0.026 & 0.046 & 0.046 & 0.053 & 0.034 & 0.122 & 0.070 \\
$\rho_{xx: 2 \textrm{K}} (\mu \Omega \textrm{cm}$) & 7.5 & 20 & 25 & 60 & 82 & 102 & 258 \\ 
$\rho_{xx: 80 \textrm{K}} / \rho_{xx: 2 \textrm{K}}$ & 7.9 & 8.0 & 6.2 & 3.0 & 2.0 & $\sim$ 3 & 1.6 \\ 
\hline
\end{tabular}
\caption{List of main characteristics of various MnSi samples from resistivity measurements: ordering temperature $T_C$, resistive coefficient $A$, resistivity $\rho_{xx: 2 \textrm{K}}$ at 2\,K, resistivity ratio $\rho_{xx: 80 \textrm{K}} / \rho_{xx: 2 \textrm{K}}$; for details see text \cite{note2}.}
\label{tab:samples}
\end{table}

For bulk material, a kink in the resistivity reflects a magnetic transition with an ordering temperature $T_C = 29$\,K, consistent with previous reports \cite{pfleiderer2010,kadowaki1982}. For thin film material, with the measurement of the resistivity we again reproduce the behavior reported in literature \cite{engelke2012,li2013,yokouchi2014,meynell2014b,yokouchi2015}. The residual resistivity is significantly larger than that of bulk material, be it that the resistivity ratio is also slightly larger, indicating a reasonable crystalline quality of the thin film material. Overall, the values we obtain for these quantities for our samples are comparable to those reported in literature (see Tab.\,\ref{tab:comp}), implying an overall similar film quality. As noted before, the magnetic transition temperature is enhanced as compared to bulk, reflecting the tensile strain/effective negative pressure.

The next processing step, nanostructuring the thin films, does not affect the overall appearance of the resistivity $\rho_{xx}$. All nanostructured samples exhibit a kink in the resistivity at $T_C \sim 44 - 48$\,K (for the 230\,nm sample, see Ref. \cite{note2}), denoting the magnetic transition, as in thin film material. As well, below $T_C$ the resistivity evolves as $\rho_{xx} = \rho_{0} + A T^2$. We have included the values $A$ from fits to the data below $T_C$ in Tab. \ref{tab:samples}, yielding values slightly larger than for bulk material for all samples, and consistent with the thin film result. 

Surprisingly, nanostructuring the thin film material affects a characteristic parameter of the electronic transport, {\it i.e.}, the residual resistivity increases with smaller nanostructure size. While down to a structure width of 300\,nm the resistivities are comparatively large, but still in a metallic range, for our smallest structures we find (residual) resistivities in the range of a few hundred $\mu \Omega \textrm{cm}$. Correspondingly, the resistivity ratio decreases with nanostructure size, reflecting the increase of $\rho_{xx: 2 \textrm{K}}$.

Assuming a homogeneous current path in the sample, resistivities above the Mooji rule \cite{mooji,tsuei} ($\sim 200$\,$\mu \Omega$cm) are typically accounted for in terms of either a semi-metallic system or disorder-induced localization. Clearly, the second scenario of disorder-induced localization is not consistent with observations, as in this case (a tendency towards) a negative resistive coefficient $d \rho/d T$ should occur. As well, a transition of MnSi into an intrinsically semi-metallic state induced by nanostructuring appears unlikely, as the magnetic behavior and the character of the temperature dependent resistivity is not affected by the structuring.

But then, in order to account for the large resistivity values in our smallest structure MnSi, either the concept of a homogeneous current path in our nanostructure is not fulfilled, or we are dealing with interfacial effects of an unknown nature in a system consisting of Si and the correlated electron material MnSi.

\subsection{Magnetoresistance}
\label{magnetoresistance}

\begin{figure}
\centering
		\includegraphics[width=1 \columnwidth]{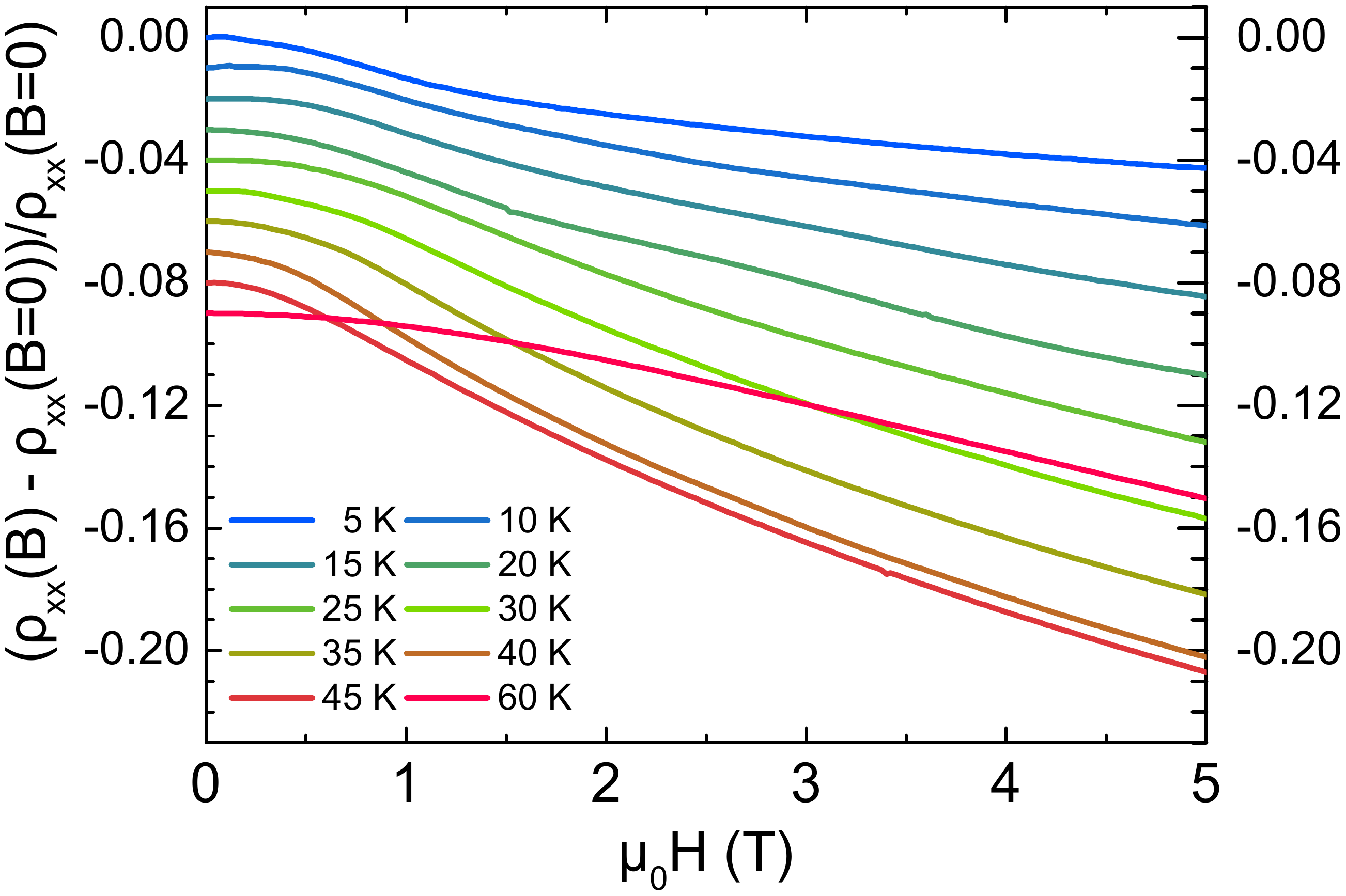}
	\caption{(Color online) Magnetoresistivity of a 10\,$\mu$m MnSi thin film structure for a magnetic field $B \perp I$ up to 5 T and different temperatures; data shifted for clarity, for details see text.}
	\label{magnetores}
\end{figure}

To further characterize our nanostructure samples MnSi we have studied the magnetotransport properties. As an example, in Fig. \ref{magnetores} we plot the transverse magnetoresistivity $(\rho_{xx}(B) - \rho_{xx}(B=0))/\rho_{xx}(B=0)$ of a MnSi structure (10\,$\mu$m current path width). Above $T_C$, the magnetoresistivity is negative and evolves $\propto B^2$, analogous to the findings in single crystalline material \cite{sakakibara1982,kadowaki1982} and in agreement with previous thin film studies. Such behavior has been interpreted in terms of spin fluctuation theory \cite{ueda1976}.

As the temperature is lowered below $T_C$, the shape of the magnetoresistivity qualitatively transforms into an inverted S-shape. Moreover, in the temperature range $\sim 10$\,K a small kink in $\rho(B)$ appears near 1.2\,T. Such a feature has been seen in previous measurements on MnSi thin film \cite{menzel2013} and been associated to the phase transition from the magnetically ordered into the ferromagnetically polarized state at $B_{C}$ \cite{kadowaki1982}. Weak additional kinks in the magnetoresistivity reported previously below $B_{C}$ and tentatively also associated to skyrmion phase formation \cite{meynell2014b} were not observed in our studies. 

\begin{figure}
\centering
		\includegraphics[width=1 \columnwidth]{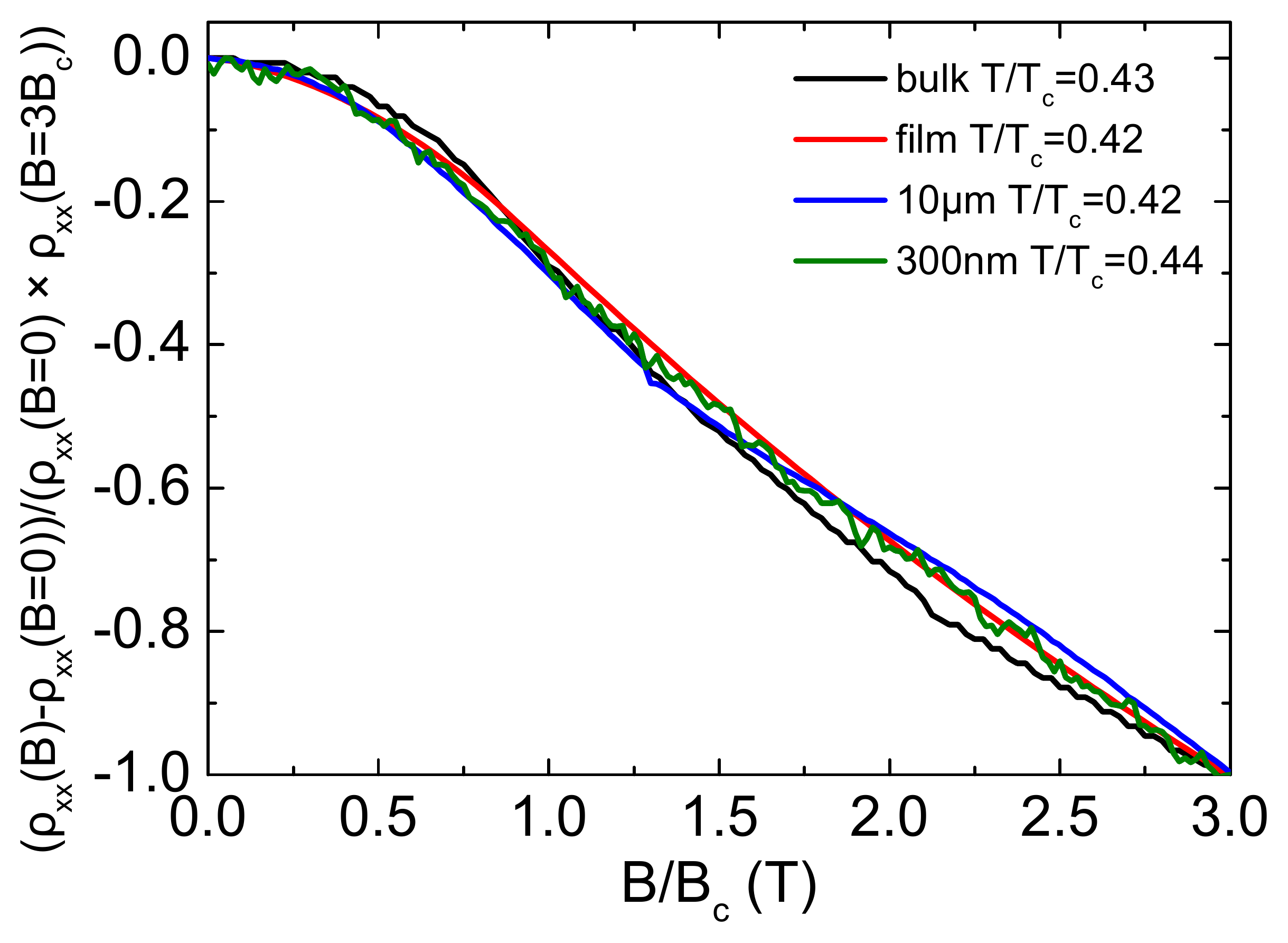}
	\caption{Comparision of the magnetoresistivity of different nanostructured samples MnSi for a magnetic field $B \perp I$ at a reduced temperature $T/T_C \sim 0.43$ as function of the reduced field $B/B_{C}$ with the magnetoresistive effect scaled to the value at $B = 3 \times B_{C}$; for details see text.}
	\label{magnetorescomp}
\end{figure}

The magnetoresistive behavior reported here for the 10\,$\mu$m structure is similarily seen for all the other structures. This is illustrated in Fig. \ref{magnetorescomp}, where we plot the magnetoresistivity of different nanostructured samples MnSi at a reduced temperature $T/T_C \sim 0.43$ as function of the reduced magnetic field $B/B_{C}$, and scaled to a value of the magnetoresistivity MR $:= -1$ at $3 \times B_C$. Clearly, for all samples there is a close resemblance in the field evolution, and the features possibly denoting phase transitions coincide.

Altogether, in terms of the resistivity and magnetoresistivity, with our measurements on nanostructured MnSi samples we qualitatively and quantitatively reproduce the findings previously reported on thin film material. Correspondingly, we can safely assume that the physical behavior observed for our samples may be compared to that reported previously for thin film material. Conversely, we still need to explain the enhancement of the residual resistivity for our nanostructured samples.

\subsection{Hall effect}
\label{hallsection}

Finally, we have performed Hall effect measurements on our various nanostructured samples MnSi using an out-of-plane magnetic field $B$, the current $I$ along the structure, and the Hall voltage $U_H$ measured $\perp I$ and $\perp B$ (see Fig. \ref{rem}(a)). In addition, in order to allow for a direct comparison to bulk material MnSi, we have carried out corresponding measurements using the same experimental set-up on the single crystal, for which we have reported the (magneto)resistivity. As well, we have measured the Hall effect for our thin film sample MnSi.

\begin{figure}
\centering
		\includegraphics[width=1 \columnwidth]{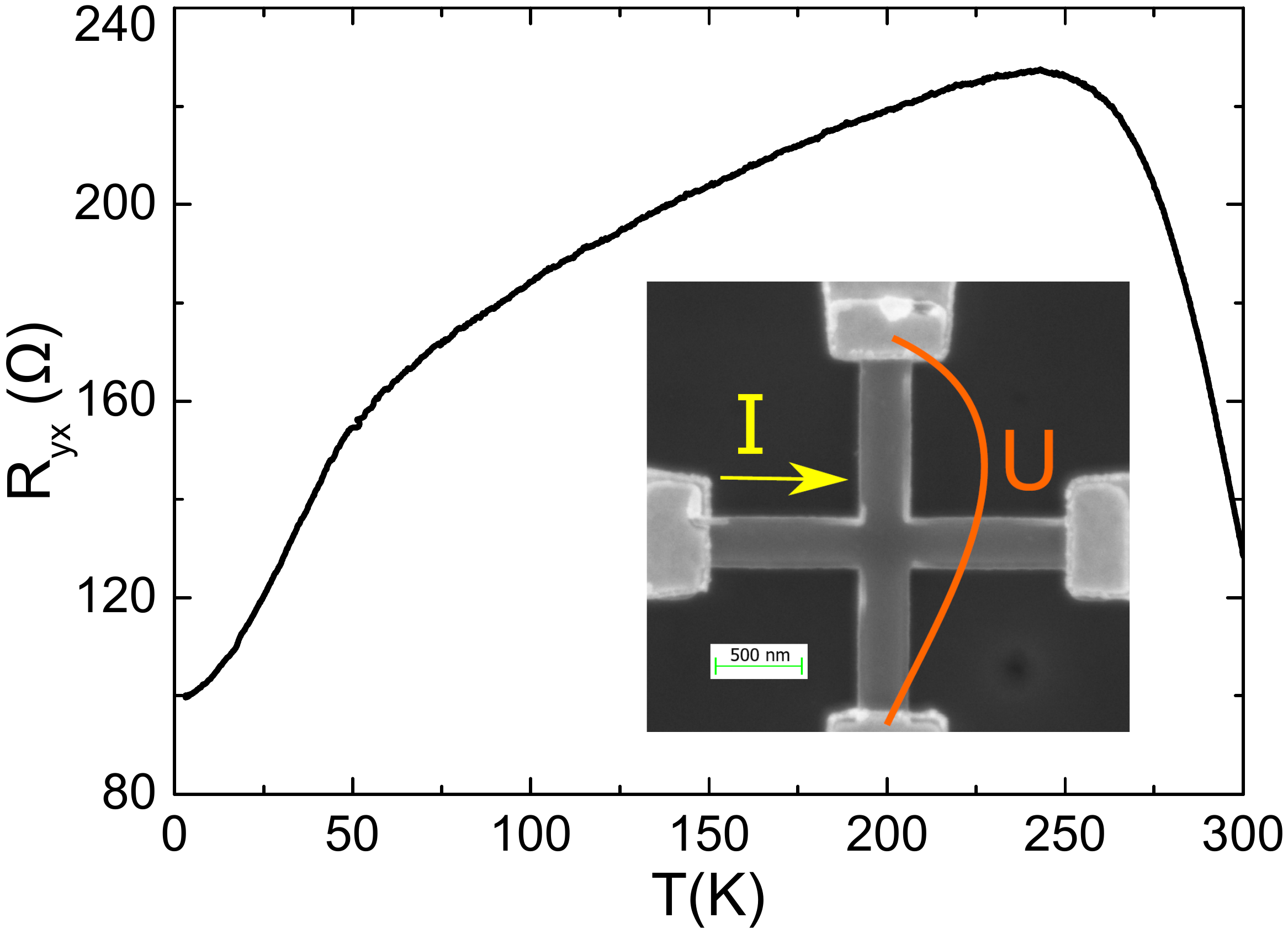}
	\caption{Absolute resistance for a MnSi cross structure (see scanning electron microscope picture in the insert) in zero magnetic field.}
	\label{cross}
\end{figure}

Surprisingly, for our Hall bar structures, in our initial experiments we find - in addition to a Hall voltage $U_H$ - a significant and sample dependent magnetoresistive contribution $U_R$. Following these observations, we have simplified our experimental geometry by producing a single Hall cross (see SEM-picture in the inset of Fig. \ref{cross}; current path width 290\,nm) and measured the cross voltage $U_{cross}$ in zero magnetic field for this structure (Fig. \ref{cross}). 

In a Hall effect configuration the experimentally measured zero-field offset and magnetoresistive contribution is caused by the residual geometrical offset of the voltage leads. For our configuration, clearly, this offset is technically minimized to a writing mismatch from the e-beam lithography ({\it i.e.}, nanometer range), and which should produce only a very small resistive signal. Still, experimentally we do observe a large temperature dependent cross voltage that clearly resembles the resistance measurements presented above. By comparing the absolute resistance scale of the single cross seen in Fig. \ref{cross} with measurements performed on a Hall bar structure with comparable current path width between two neighboring voltage taps (single cross current path width $\sim 290$\,nm, Hall bar structure width $\sim 300$\,nm and length of $\sim 2.5$\,$\mu$m, same unpatterned film origin) we see a resistive signal of the same order, even though we are now measuring the voltage drop on nominally equipotential voltage leads. 

To quantify the anomalously large cross voltage/resistance for this particular nanostructure, we can estimate the geometrical shift of the voltage leads with respect to each other that would be required to account for the observations. With the measured cross resistance $R_{cross}$ of the structure with a current path width $w = 290$\,nm, the resistivity at 5\,K, $\rho_{300 {\rm nm}} = 82 \mu \Omega$cm, and film thickness $d = 30$\,nm, we can calculate the geometrical shift $x$ as
\begin{equation}
x = \frac{R_{cross}\cdot d \cdot w}{\rho_{300 {\rm nm}}}. 
\end{equation}     
This way, we find a value $x = 1066$\,nm, which is of the order of the structure size. Clearly, such behavior is at odds with our assumption of a homogeneous metallic resistance path that we are probing. 

One might argue that the very large apparent resistance path that we report for the simple Hall cross in Fig. \ref{cross} represents a device failure, even though the temperature dependence of the resistance nicely follows the behavior expected for thin film MnSi. Therefore, we have evaluated the resistive component for our different Hall voltage structures across the equipotential contacts. From this analysis, we find that for all nanostructure samples there is a detectable sample/device-dependent zero field transverse Hall resistance: For the 290\,nm cross we find the largest resistance value, about 100\,$\Omega$, while for the other samples it varies between 0.01\,$\Omega$ (thin film) and 20\,$\Omega$ (175\,nm Hall cross), with seemingly a tendency towards larger resistance values for smaller structures. Even disregarding the more extreme values we still see a variance of the zero field transverse Hall resistance of nominal equipotential voltage taps in the order of 0.01 to 1\,$\Omega$. These resistance values need to be viewed in relation to the Hall resistance across the contacts, which is of the order of 0.05\,$\Omega$. In this situation, for a large residual resistance on equipotential contacts the signal-to-noise ratio for the determination of the Hall resistance becomes rather small. Below, we will discuss the consequences of this finding.

\begin{figure}
\centering
		\includegraphics[width=1 \columnwidth]{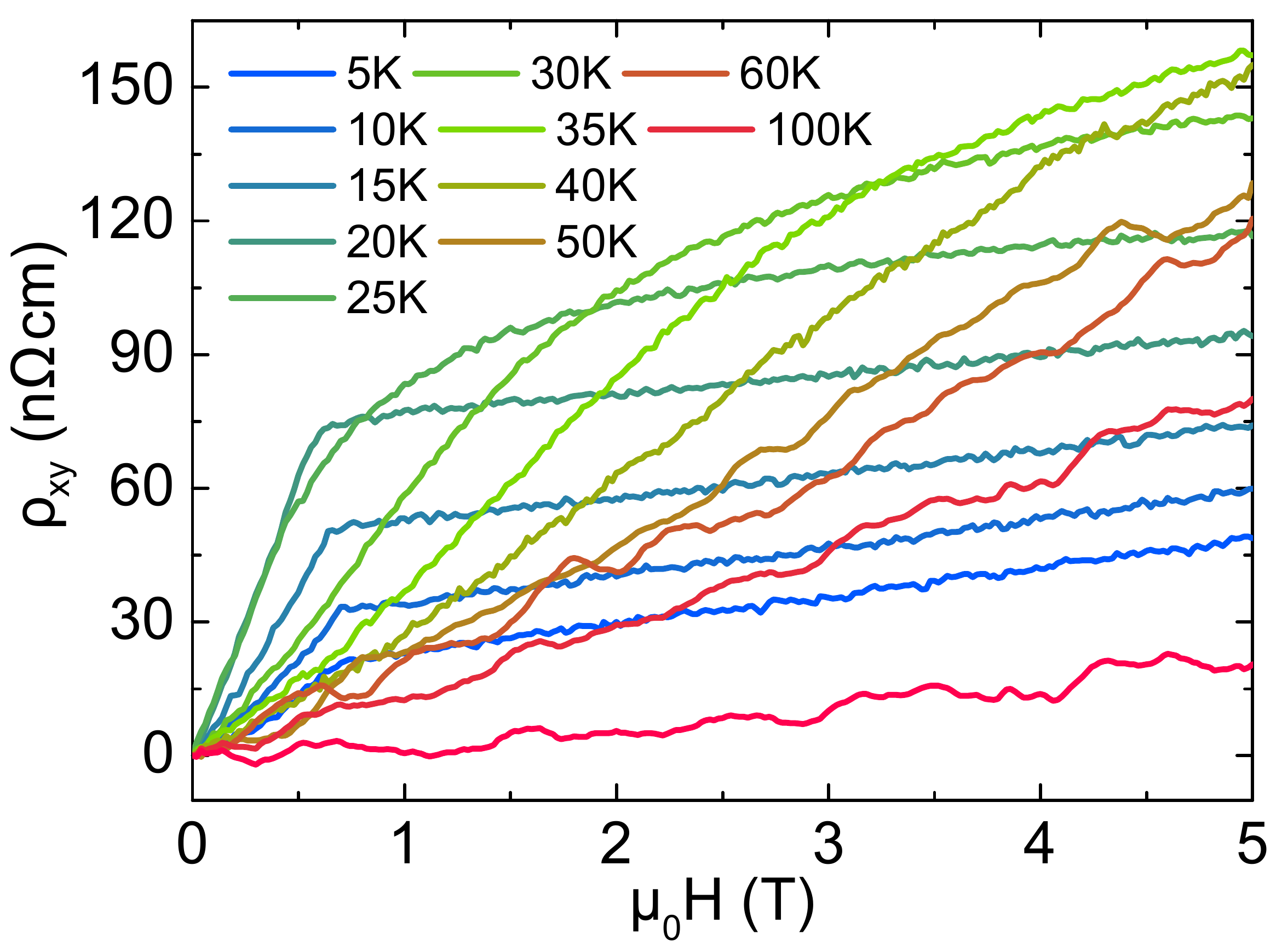}
	\caption{The anomalous Hall effect $\rho_{yx}^{A}$ of single crystalline bulk MnSi.}
	\label{hallbulk}
\end{figure}

\begin{figure}
\centering
		\includegraphics[width=1 \columnwidth]{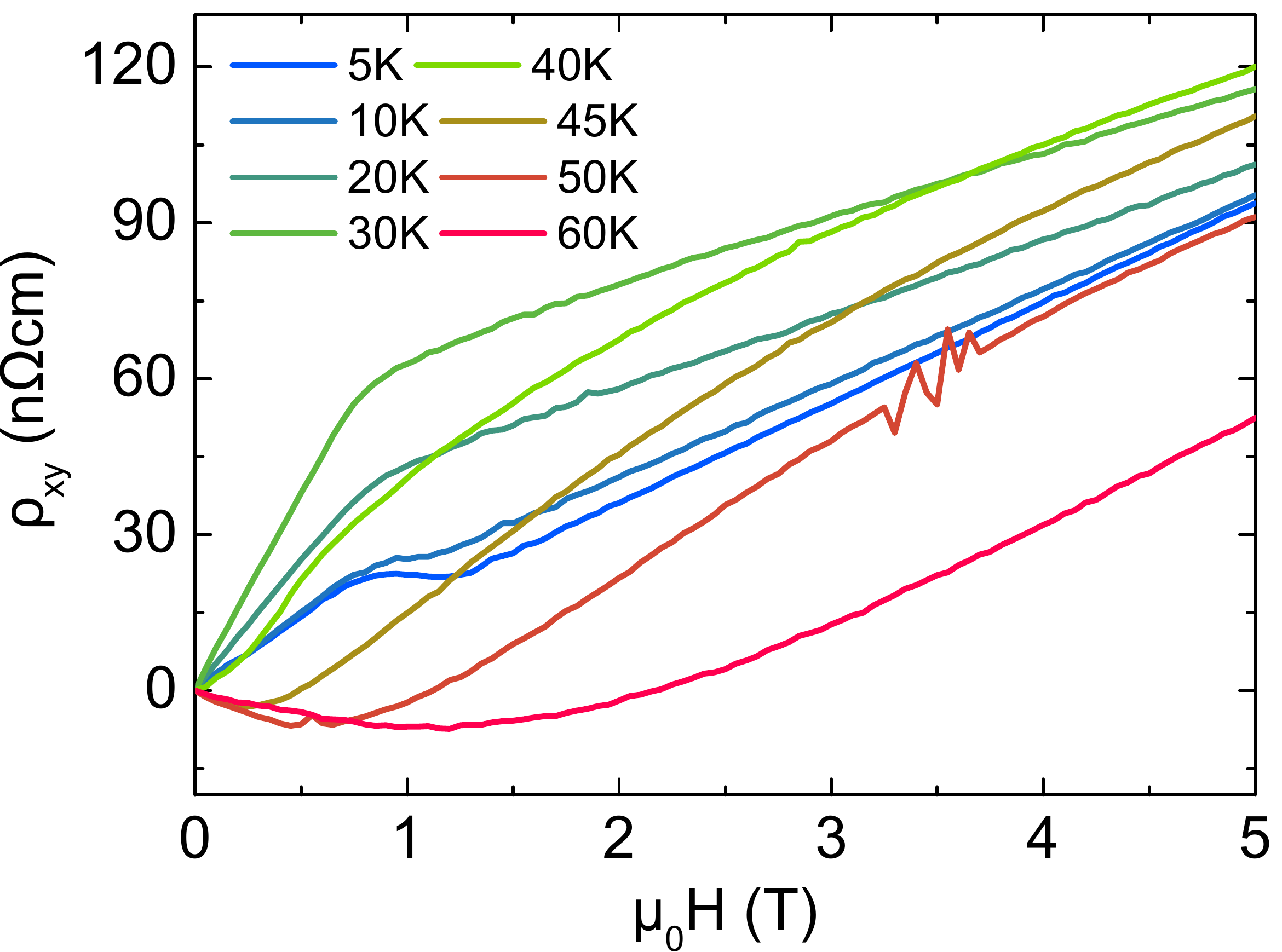}
	\caption{The anomalous Hall effect $\rho_{yx}^{A}$ of thin film MnSi (film thickness 30\,nm).}
	\label{hallfilm}
\end{figure}

\begin{figure}
\centering
		\includegraphics[width=1 \columnwidth]{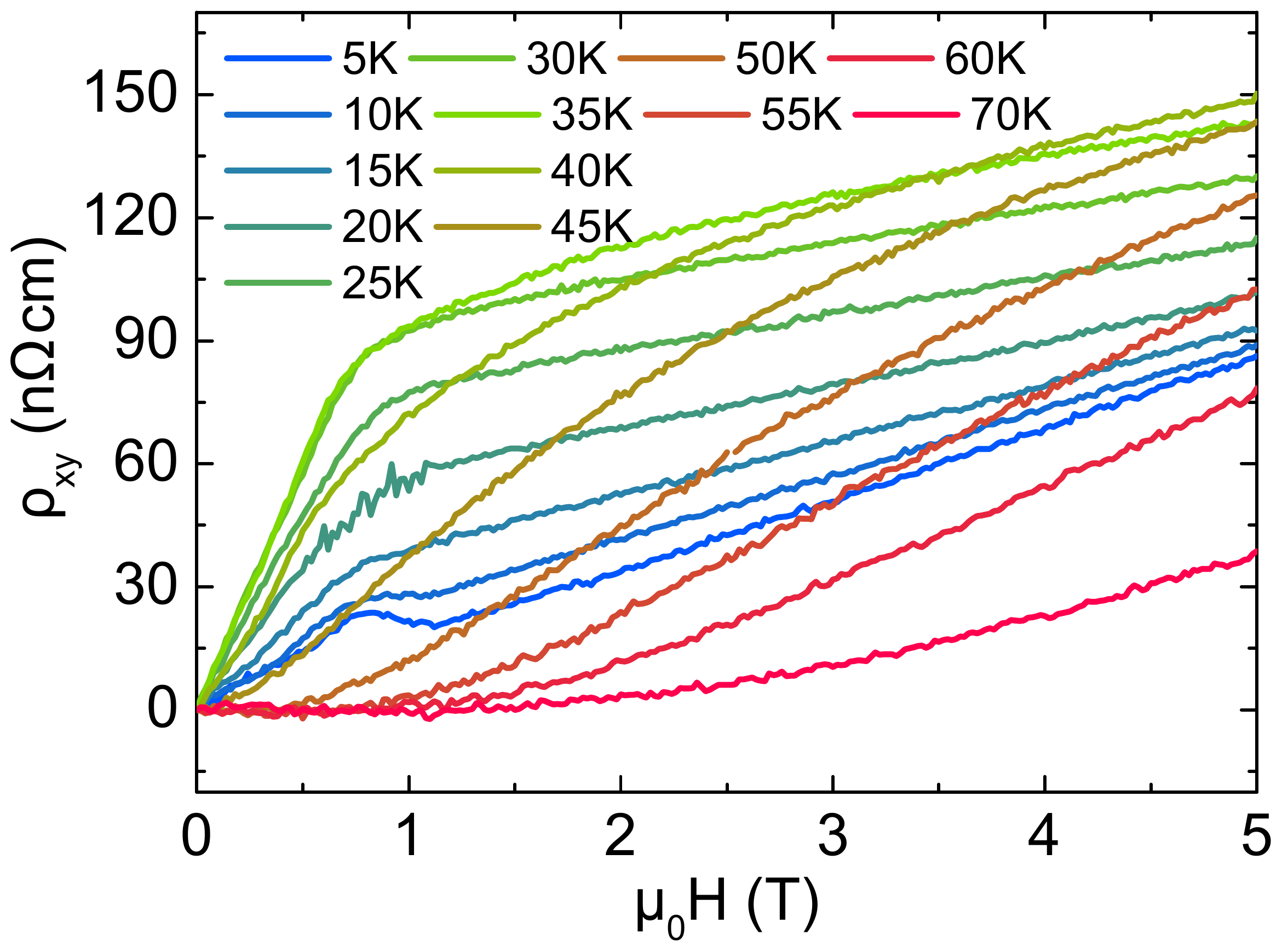}
	\caption{The anomalous Hall effect $\rho_{yx}^{A}$ of nanostructured MnSi (structure width 1\,$\mu$m).}
	\label{hall01mu}
\end{figure}

\begin{figure}
\centering
			\includegraphics[width=1 \columnwidth]{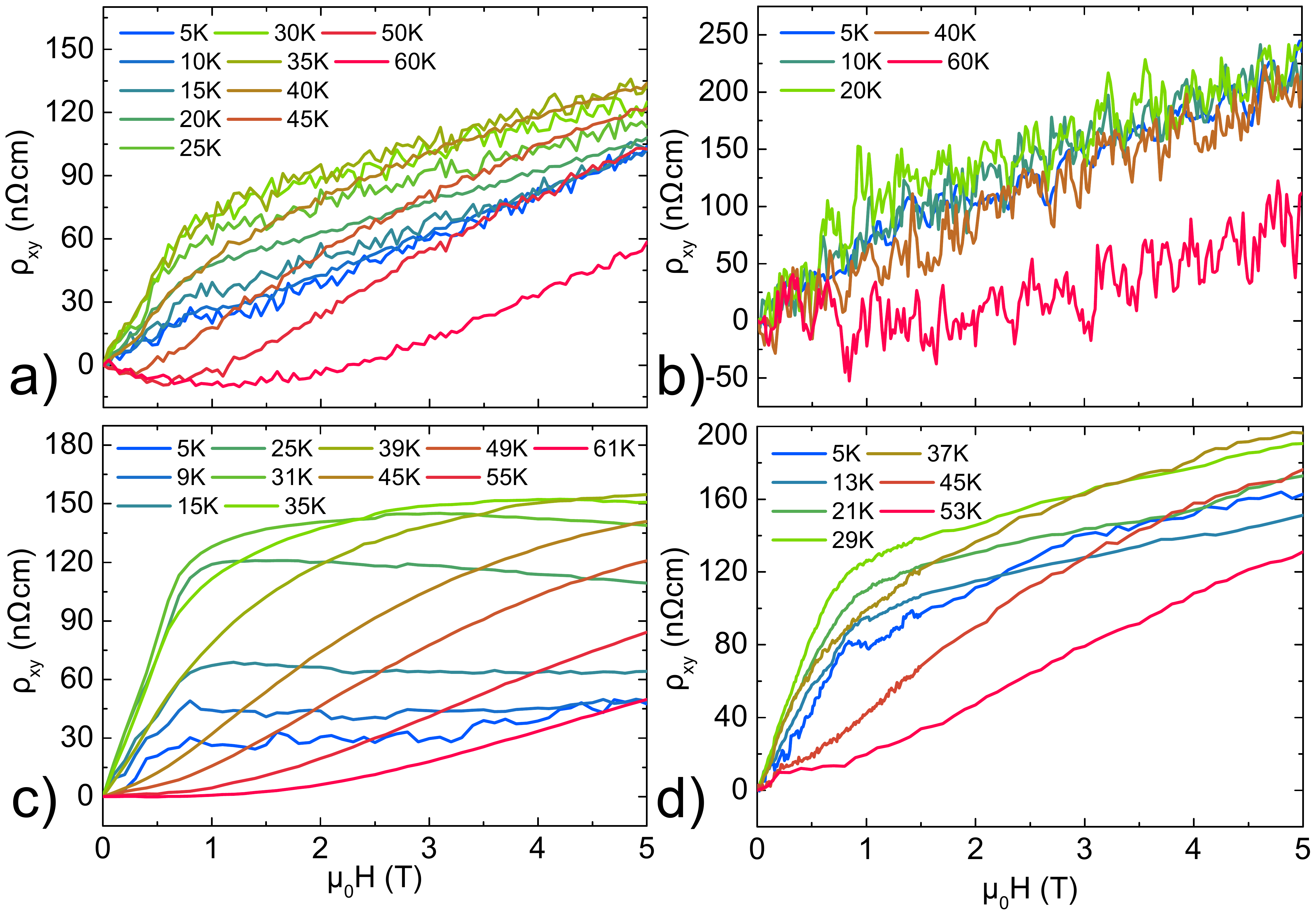}
	\caption{The anomalous Hall effect $\rho_{yx}^{A}$ of nanostructured MnSi with structure width of a) 10 $\mu$m, b) 300\,nm, c) 230\,nm and d) 200\,nm.}
	\label{hallstruct}
\end{figure}

Next, to finalize our comparative analysis of nano\-structured thin film MnSi we have determined the Hall resistivity $\rho_{yx}$ for our different samples (see Figs. \ref{hallbulk} -- \ref{hallstruct}). Because of the (magneto)resistive signal component discussed before, to extract the Hall resistivity, even for the nanostructured samples we have to determine the field-symmetric and antisymmetric signal contribution, with the latter one representing the Hall signal. 

Again, we start our discussion with the data for bulk material. With the approach from Eq. (1) to separate the Hall effect into normal and anomalous contribution, for the normal part we can write $\rho_{yx}^N = R_0 B = (n e)^{-1} B$, with the Hall coefficient $R_0$ and the carrier density $n$ of electron charges $e$. Thus, the observation of a linear-in-field behavior of $\rho_{yx}$ at high temperatures (100\,K $\gg T_C$), in agreement with the findings of Ref. \cite{neubauer2009}, allows corresponding fits to the data, with a summary of the results listed in Tab. \ref{tab:table1}. We find a quantitative difference in the absolute value of the Hall resistivity between our data and that from Ref. \cite{neubauer2009}, which translates into a carrier density for our crystal being the corresponding factor larger. It is not clear, if this difference in the value of $n$ reflects a sample dependence. We note that the room temperature resistivity of our sample is 30 \% smaller than that from Neubauer et al. \cite{neubauer2009}. If this difference arises from an inaccurate determination of the (effective) sample geometries in the one or other case, it would translate into a difference of the Hall coefficient broadly consistent with the present data. 

\begin{table}
\caption{Measured Hall coefficient $R_0$ and the resulting charge carrier density $n$ for different samples MnSi; for details see \cite{note1}.}
\begin{tabular}{lcc}
\br
\textrm{Sample}&
\textrm{R\textsubscript{0} ($\times10^{-10}$ $\Omega mT^{-1}$})&
\textrm{n $(10^{22}$\,cm$^{-3}$)}\\
\mr
Bulk & 0.8 $\pm$ 0.3 & 7.7 $\pm$ 1.9\\
Thin film & 3.4 $\pm$ 0.3 & 1.8 $\pm$ 0.2\\
Structure 1 (10 $\mu$m)  & 3.7 $\pm$ 0.3 & 1.7 $\pm$ 0.2\\
Structure 2 (1 $\mu$m)  & 3.3 $\pm$ 0.2 & 1.9 $\pm$ 0.1\\
Structure 3 (300 nm)  & 6.4 $\pm$ 1.7 & 1.0 $\pm$ 0.3\\
Structure 4 (230 nm) & 2.4 $\pm$ 0.8 & 2.6 $\pm$ 0.2\\
Structure 5 (200 nm) & 3.6 $\pm$ 0.4 & 1.7 $\pm$ 0.1\\
Structure 6 (175 nm)  & 11.8 $\pm$ 2.5 & 0.5 $\pm$ 0.1 \\
\br
\end{tabular}

\label{tab:table1}
\end{table}

After subtraction of the normal Hall contribution we obtain the anomalous contribution, {\it i.e.}, $\rho_{yx}^{A} = \rho_{yx} - \rho_{yx}^N (300 {\rm K})$ for bulk material, which we plot in Fig. \ref{hallbulk}. Qualitatively and -- as discussed -- semiquantitatively our data reproduce the behavior reported in Ref. \cite{neubauer2009}. Overall, our measured anomalous Hall resistivity is a factor of two smaller than that reported in Ref. \cite{neubauer2009}, but otherwise the temperature and field evolution of $\rho_{yx}^{A}$ is the same. The anomalous Hall contribution is positive, its field dependence essentially reflects that of the magnetization, with a prefactor that increases with temperature up to about $T_C$, in good agreement with Ref. \cite{neubauer2009}.

Next, we have performed Hall effect measurements on our thin film sample. The overall appearance roughly resembles the single crystal data, but with some differences in detail. As before, at high temperatures we find a linear-in-field behavior that we use to extract the carrier density. We find it to be about a factor of four smaller than for our single crystal, and a factor of two smaller than the crystal value reported in Ref. \cite{neubauer2009} (see Tab. \ref{tab:table1}).

Again, the anomalous contribution we have extracted following the same procedure as for the single crystal. The overall evolution of the anomalous contribution is similar to that of the crystal: The field dependence shows basically a magnetization behavior, and the overall signal size increases with increasing temperature up to about 50\,K, {\it i.e.}, $T_C$ (Fig. \ref{hallfilm}). Only, compared to the single crystal data, the magnitude of the anomalous contribution again is somewhat larger, with an overall signal change of about 100\,n$\Omega$cm as compared to 70\,n$\Omega$cm for the single crystal.

Finally, we have carried out Hall effect measurements on our nanostructured samples MnSi, with the main results summarized in the Figs. \ref{hall01mu} and \ref{hallstruct}. Again, the structures exhibit a behavior which overall is qualitatively similar to that of the crystal, but with some differences in detail. First, from the high-temperature behavior we extract the carrier densities as before (Tab. \ref{tab:table1}). Here, the carrier density determined this way appears to change with structure size, with for the smallest structure being an order of magnitude smaller than for single crystalline material.

After subtraction of the normal contribution we obtain the anomalous contribution. As before, qualitatively it resembles that of the crystal and the thin film, producing a positive signal with a magnetization-like field dependence. As well, for increasing temperature up to $\sim 50$\,K, the signal amplitude increases. As before, the amplitude is enhanced as compared to the crystal, and also slightly larger than for the thin film (about 120\,n$\Omega$cm).

We have also attempted to measure the Hall effect for our smallest structure (path width 175\,nm). Here, however, in the cross voltage $U_{cross}$ the magnetoresistive contribution to the signal was much larger than the Hall part. Therefore, because of the low signal-to-noise ratio mentioned above it was impossible to clearly resolve the field dependence of the Hall voltage. We note that already for the structure with path width 300\,nm an enhanced noise level is observable in the Hall effect data (see Fig. \ref{hallstruct}), and which stems from the same difficulty.

\section{Discussion}

Summarizing our observations so far, we have performed a thorough and comparative study on single-crystalline, thin film and nanostructured MnSi. Qualitatively and quantitatively, we reproduce the findings on resistivity and magnetotransport previously reported for crystalline and thin film samples. Moreover, we have produced a consistent set of Hall effect data for our set of samples, which globally exhibit a similar behavior and are qualitatively in agreement with some of the prior reports \cite{karhu2010,karhu2011,karhu2012,geisler2012,wilson2012,engelke2012,li2013,suzuki2013,menzel2013,wilson2013,yokouchi2014,wilson2014,engelke2014,meynell2014a,meynell2014b,yokouchi2015,lancaster2016}, thus allowing us to discuss our observations in comparison to and in the context of those studies.

The essential findings from our comparative study on thin film and nanostructured MnSi to be discussed in detail are thus:
\begin{itemize}
	\item the measured (residual) resistivity increases with decreasing structure size;
	\item for the smallest structures the measured resistivity values reach into the ''non-metallic'' range, while the overall metallic behavior ($d \rho /d T > 0$) of and magnetic transition signatures in the resistivity are preserved;
	\item the magnetoresistivity is not affected in its appearance by the structure size;
	\item even dedicated Hall bar structures produce a resistive signal in a Hall geometry;
	\item the carrier density derived from the thin film/nanostructure measurements appears to be much smaller than for single crystalline material, with some indications of a structure size dependence.
\end{itemize}

Some of these observations can be accounted for in rather simple terms. As pointed out, the observation of a large resistive contribution for the measurement of the Hall cross in Fig. \ref{cross} and the measured magnetoresistive component for the Hall structures requires to drop the assumption of a homogeneous current path for all samples. Instead, if we assume that the current follows a (more-or-less) percolative path through the sample, one could easily obtain a significant resistive signal even in Hall cross geometry, {\it viz.}, on nominally equipotential points in the structure.

The concept of a percolative current path in thin film/nanostructured MnSi would account for the observed increase of the resistivity with decreasing structure size. Conceptually, a percolative current path in a metallic system reflects that in the material there are spatial regions of high and low conductivity. For a thin film, the regions of high conductivity will dominantly carry the current and can be understood as a network of parallel/series resistors. An extended thin film corresponds to a very large network of resistors, and their collective behavior will be similar to the case of a homogeneously distributed current, {\it i.e.}, the film and bulk behavior are similar. 

In contrast, when we limit the size of the system by nanostructuring, the extension of the network is gradually reduced. In result, this will typically lead to an effectively increased current path. For sufficiently long current paths it may even lead to measured resistances of the structures, which - if transformed into resistivities by using the nominal geometrical dimensions of the structures - produce $\rho$-values that exceed typical metallic values without loosing the metallic character $d \rho /d T > 0$. As well, with this scenario a pronounced device dependence might be expected, as structural defects such a grain boundaries - if they span the whole device - might have a significant effect on the measured resistance. Conversely, a percolative current path would not show up in the magnetoresistivity, if we use a scaled representation as performed in Fig. \ref{magnetores}. 

The concept of a percolative current path in MnSi films/structures might thus explain certain findings reported here. However, reconciling it with the metallicity of the MnSi films appears not to be easy. If we assume that the current in thin films MnSi follows a percolative path, we imply that we have produced significant spatial conductivity variations in a metallic system. One may argue that the type of island growth that controls the morphology of MnSi is responsible for this, effectively producing many grain boundaries, where local strain etc. will reduce conductivity. Consistent with this concept is the observation that the size of the islands in MnSi is typically of the order of a few 10 to 100\,nm (see shade variations in the inset of Fig. \ref{afmzoom}, which track the island growth morphology). With our smallest nanostructures MnSi we would thus probe single island regimes, where we might expect large effects from percolative resistivity. Only, based on experience with common metallic films, the naive expectation would be that local conductivity differences even across different islands would be of the order of $\sim 10$\,\%, and it is not obvious that this would be sufficient to produce a percolative resistance path.

We may speculate that part of this issue relates to the tensile strain in the thin films MnSi. The nanostructuring process might lead to a lateral relaxation of the MnSi thin film. This may increase the strain at the grain boundaries and, thus, enhance the percolative resistive effect. If this simple picture of structural relaxation affecting the electronic transport properties of nanostructured MnSi films carries some truth, it would imply that in order to produce and study nanostructured material it requires improvements to the sample quality by growing unstrained epitaxial films.  

Regarding the Hall effect, the idea of thin film MnSi being electronically inhomogeneous might provide explanations for some experimental observations. For instance, the sample-to-sample variations of the anomalous Hall contribution, as summarized in Table \ref{tab:comp}, might simply reflect that because of an inhomogeneous current path only certain parts of the samples are probed by the Hall effect. As these are different from sample to sample, they produce corresponding differences in the anomalous Hall effect. If this would be true, of course, in the future the search for skyrmions by means of the (topological) Hall effect will need to verify that with the Hall effect the spatial regions containing skyrmions are probed. As well, if working on nanostructure samples, to classify these it will be important to verify that in Hall geometry there is only a small zero-field cross voltage ({\it i.e.}, a relatively homogeneous current path can be assumed). 

As well, the argument of an electronical inhomogeneity would suggest that it varies on a spatial range of the order of the MnSi islands, {\it i.e.}, a few ten nanometers. Most likely, this length scale is also an upper limit for the electronic mean free path, leading to a situation where the mean free path is not significantly larger than the diameter of a skyrmion. In that case, it is not evident if the electrons at all can ''see'' a skyrmion, as they might undergo scattering before traversing the skyrmion. Again, this might imply that in thin films MnSi it would be difficult to see skyrmionic phases by means of the (topological) Hall effect.

Finally, the apparent reduction of the carrier density with reduced structure size needs to be discussed. Clearly, our experimental observation can not reflect a true carrier density reduction in our nanostructure samples by an order of magnitude compared to bulk material, as this should significantly affect the magnetic properties. Given that $T_C$ is constant for the thin film/nanostructure samples, the magnetically ordered areas in the samples will have typical metallic carrier densities. But then, the reduction of the carrier density - which corresponds to an apparent increase of the Hall voltage - must be an artifact. Here, we might consider if the concept of a percolative resistance accounts for our observations. 

In the most basic approach (neglecting band structure etc.), the Hall voltage over a sample is calculated as
\begin{equation}
U_H = \frac{R_H \cdot I \cdot B}{d}, 
\end{equation}
with the Hall coefficient $R_H$, the current through the sample $I$, the externally applied magnetic field $B$ and the sample thickness $d$. As a note of caution, we stress that this equation is derived based on the concept of a homogeneous current path, and which possibly is not fulfilled in all our cases to good approximation. Still, using the equation as a starting point, in a comparative study on nanostructures produced from the same or similar thin film samples, an artificial enhancement of $U_H$ can not result from a variation of the film thickness $d$. One might argue that with the Hall effect we are predominantly probing sample areas with low carrier densities (large local $R_H$), but in the metallic environment of the surrounding area a short-cutting of such effects would be expected. It remains the possibility of an artifical enhancement of the local current/current density that might produce an increase of $U_H$ measured across the Hall cross area. Such a scenario might possibly arise from the concept of a percolative resistance path sketched above, as with a more narrow current path local currents might be larger. To definitely answer if this is the case for our nanostructure samples MnSi, however, it would require better knowledge on the local structural and electronic ''morphology'' of our samples. 

Altogether, for thin film/nanostructured MnSi a picture emerges, where thin film material behaves differently from single crystalline specimens. Our work leads us to conclude that there is a possibility of an electronic inhomogeneity in samples of MnSi. This concept as such would be highly unusual for simple metallic systems, but has been discussed for correlated electron systems related to MnSi. In recent years, it has been demonstrated that disorder effects are very much enhanced for correlated electron materials, and which have been discussed for instance within the context of Griffiths phases, which effectively consider electronic inhomogeneities in correlated metals \cite{castro1998,otop2005,brando2016}. Here, a line of thought might be that the effect of disorder induced strain in the films is enhanced by the correlations, which in in turn might produce an electronic inhomogeneity.

In the context of skyrmionic physics, our work has also some consequences. Notably, the observation of local objects such as skyrmions by rather delocalized measurment techniques such electronic transport/Hall effect might be difficult in the presence of structural inhomogeneities. Conversely, the role of structural inhomogeneities for the creation or destruction of skyrmions has also not been studied in detail. Thus, our work highlights the relevance of understanding the role of structural disorder for skyrmionic systems, especially in terms of testing skyrmions electronically.

\ack
We gratefully acknowledge support by the Braunschweig International Graduate School of Metrology B-IGSM and the DFG Research Training Group GrK1952/1 ''Metrology for Complex Nanosystems''.

\appendix


\begin{thebibliography}{00}
\bibitem{note2} The resistivity of the 230\,nm sample was measured with a current density of $2\cdot10^{10}\,$A/m$^2$ which is 10$^{3}$ $-$ 10$^{4}$ times higher then the current density used for the other measurements. Consequently the $\sim$ 5\,K shift of the transition temperature is most likely explained via a heating effect in the structure.
\end{thebibliography}
\end{document}